# A Systemic IoT-Fog-Cloud Architecture for Big-Data Analytics and Cyber Security Systems: A Review of Fog Computing

Nour Moustafa
School of Engineering and Information Technology
University of New South Wales-Canberra @ ADFA
Canberra, Australia
*nour.moustafa@unsw.edu.au*

*Abstract*—With the rapid growth of the Internet of Things (IoT), current Cloud systems face various drawbacks such as lack of mobility support, location-awareness, geo-distribution, high latency, as well as cyber threats. Fog/Edge computing has been proposed for addressing some of the drawbacks, as it enables computing resources at the network's edges and it locally offers big-data analytics rather than transmitting them to the Cloud. The Fog is defined as a Cloud-like system having similar functions, including software-, platform- and infrastructure-as services. The deployment of Fog applications faces various security issues related to virtualisation, network monitoring, data protection and attack detection. This paper proposes a systemic IoT-Fog-Cloud architecture that clarifies the interactions between the three layers of IoT, Fog and Cloud for effectively implementing big-data analytics and cyber security applications. It also reviews security challenges, solutions and future research directions in the architecture.

## I. Introduction

The Internet of Things (IoT) has emerged to digitalise our daily tasks in various systems, for example, smart homes, smart cities, smart factories, smart grids and smart healthcare [1]. Since Cloud systems offer high computational infrastructure, power, bandwidth, software, platforms and storage, IoT applications integrate with Cloud systems across network systems [2], [3]. IoT networks include the communications of sensors, actuators and services, which require high computing resources for executing big-data analytics and cyber security applications. They still suffer from the drawbacks of scalability and operability, where heterogeneous data sources are collected and analysed from the three layers of IoT, Fog and Cloud systems [1], [4], [5].

Cloud systems, in forms of software, platforms and infrastructure, would address the challenges of scalability and operability by providing services to users and organisations. However, Cloud systems suffer from lack of mobility support, latency, location-awareness and geo-distribution [1], [6]. The Fog/Edge paradigms have been proposed to tackle the demerits of Cloud systems and enable big-data analytics at the network's edge [4]. The term 'Fog computing' was coined by the OpenFog Consortium [1], [5], which is an architecture that extends the main functions of the Cloud to provide services at the edge of a network, and is an extremely virtualised architecture of the resource pool. The Fog is a decentralised infrastructure, where data is logged and analysed between the clients and Cloud data centers. It is well-located to apply real-time and big-data analysis techniques, which considerably supports distributed data management systems [1], [3], [4], [6].

Current research studies [4]–[9] proposed that the Fog technology will be designed in the future to offer an enhanced and trustworthy architecture for handling the ever increase of interconnected appliances and services. The authors in [1], [3], [4], [6], [10], [11] suggested different methods for deploying security solutions, involving encryption, access control, firewall, authentication and intrusion detection and prevention systems, at the Fog layer. Since the Fog depends on distributed architectures which connect IoT and Cloud systems, Advanced Persistent Threats (APT) [12] could exploit Fog appliances and services if security systems are not well-designed to effectively monitor and protect the Fog nodes [1], [5].

Azam et al. [13] developed a technique for connecting a smart communication and pre-processing data module in Cloud-IoT networks. The technique integrated a smart gateway with a Fog computing technique to reduce the computation overhead at the Cloud side. Alrawais et al. [14] proposed a fog computing scheme to handle the authentication issues in IoT networks. The Fog computing device acts as a gateway of IoT devices for allocating the certificate revocation. Almadhor [15] used a Fog computing paradigm to secure Cloud-IoT platforms. Yassen et al. [16] utilised some fog computing capabilities to develop an intrusion detection system for recognising cyber-attacks in wireless sensor networks. Dsouza et al. [17] proposed a policy-based management to protect collaboration and interoperability between various customer requirements in the Fog nodes. In [18], the authors proposed a physical security framework for integrating the functions of IoT, Fog and Cloud systems. Sandhu et al. [19] proposed a framework to identify malicious activities from network edges.

In this paper, a systemic IoT-Fog-Cloud architecture is proposed for improving the execution of big-data analytics and cyber security applications. Security threats, challenges, existing security solutions, and future research directions in the Fog paradigm, is also discussed. The description of the Fog architecture is described in Section II. Section III explains security challenges and threats in the Fog. Security challenges and future directions of research are introduced in Section IV. Finally, the paper is summarised in Section V.

## II. Fog Computing Systems

### A. Description of Fog

The Fog paradigm was initially proposed by Cisco to become an extension architecture of Cloud systems that provide computation, storage and communication services between Cloud servers and client systems [1], [5], [10]. It enables computations and data processing at the network edge. This means that the Fog is a complementing layer of Cloud systems, which offers the design of a distributed architecture. The architecture can handle heterogeneous data sources of IoT wireless access networks. Big-data analytics can be implemented at the network edges faster than the centralised Cloud systems [1], [17].

The OpenFog Consortium started in 2016 for designing standardised open Fog computing frameworks [20]. For instance, an Open-Machine-to-Machine (OpenM2M) framework was suggested for linking the Fog and IoT devices and services [21]. In the framework, Fog nodes were deployed at edge infrastructures with several M2M applications. In [22], another Fog architecture was proposed, where a set of application interfaces were designed for enabling virtual machines to gain access for gathering information at Fog nodes.

Sang et al. [23] proposed a Fog framework, which is a context-aware infrastructure. The framework supports different edge technologies, including Wi-Fi, LTE and Bluetooth capabilities, which support Software Defined Networks (SDN) and virtualisation tools. It is also suggested to deploy Airborne Fog systems, where air devices like drones can perform as Fog nodes for facilitating various applications and services to end-users [24].

### B. Characteristics of Fog

Fog computing is relatively similar to Mobile-Edge Computing (MEC) and Mobile Cloud Computing (MCC) [4], [25], [26]. The MEC concentrates on Fog servers such as cloudlets that implement at the edge of mobile networks [26], whilst the MCC is an infrastructure in which both data processing and storage are executed outside of the mobile appliances [25]. The Fog has several properties that allow its integration with IoT and Cloud systems [4], [25], [26], as listed in the following:

- It locates at the network's edge and handles location awareness and low latency, as Fog nodes offer a localisation (i.e., a single hop from the device to fog node) and support end-points with rich services at the edge of a network;
- It enables dense and sparse geographical distribution, where the Fog services and application require distributed deployments;

TABLE I
SERVICES PROVIDED BY FOG/CLOUD SYSTEMS

| Fog/Cloud services | Description |
|---|---|
| SaaS | offers to a user or organisation on-demand applications and software services via a cloud infrastructure, excluding the cost of buying and maintaining these applications. Currently, Google, Amazon and Salesforce companies are the dominators of cloud service |
| PaaS | delivers to a user or organisation an application development and host client applications using libraries, services, and tools, which are supported by a PaaS provider's infrastructure |
| IaaS | offers storage, processing units, network capabilities, and other fundamental computing resources via virtual machines (VMs) to service subscribers |

- It can use large-scale sensor networks to monitor Cloud and IoT systems;
- It has a large number of nodes for demonstrating its capability of large-scale geographical distribution;
- It facilitates the mobility use which assists Fog's users to access information for improving the quality of services;
- It enables real-time interaction for handling important Fog applications;
- It supports the M2M wireless connectivity that consumes low power for supporting scalability and mobility;
- It handles different dynamic and heterogeneous sources at various levels of the network hierarchy;
- It provides flexible, inexpensive and portable deployment of hardware and software; and
- It can easily integrate IoT and Cloud applications for online big-data analytics.

### C. Systemic Architecture of IoT-Fog-Cloud

Fog computing is mainly a virtualisation technology that offers storage, computing and communication services between end devices and Cloud data centers [24], [27]. In Figure 1, a systemic architecture is proposed to show the connections of IoT, Fog and Cloud layers. An example of integrating IoT smart cities and smart factories, along with the Fog and Cloud elements, is presented. A set of IoT devices and sensors, such as green gas IoT and industrial IoT actuators, is connected to Message Queuing Telemetry Transport (MQTT) gateways to publish and subscribe to various topics, such as measuring temperature and humidity. As, in the near future, smart cities could be linked with smart factories to measure green gas emissions via IoT hubs. Therefore, it is expected that message services between various topics will be available to serve the community.

This architecture allows monitoring, filtering, inspecting, aggregating and exchanging data, resulting in saving time and computation resources for deploying and running bigdata analytics and cyber security applications [1]. Fog offers Software-as-a-Service (SaaS), Platform-as-a-Service (PaaS), Infrastructure-as-a-Service (IaaS) like cloud systems, as defined in Table I, to end-user appliances [3], [4]. In the Fog, PaaS or SaaS, the Cisco DSX was designed to establish a bridge between SaaS and different IoT devices for managing applications. This enables processing big-data at the Fog and Cloud layers for improving the computational resources of bigdata analytics and cyber security applications such as firewalls, intrusion detection and prevention systems and access control systems [1], [6].

Although the distributed architecture of the Fog can improve the computational resources of big-data analytics and network edge infrastructures, such as routers, access points, set-top-boxes and switches, should have high capabilities of CPU and GPU processors and storage [1], [3], [6]. Such infrastructures can offer computing resources as services near to customers, named fog nodes. Edge devices are considered as fog nodes, as they have computing, storage and network communications. The nodes are connected by a master-slave architecture, clustering or Peer-to-Peer networks [1], [4], such as the Cloudlet [6].

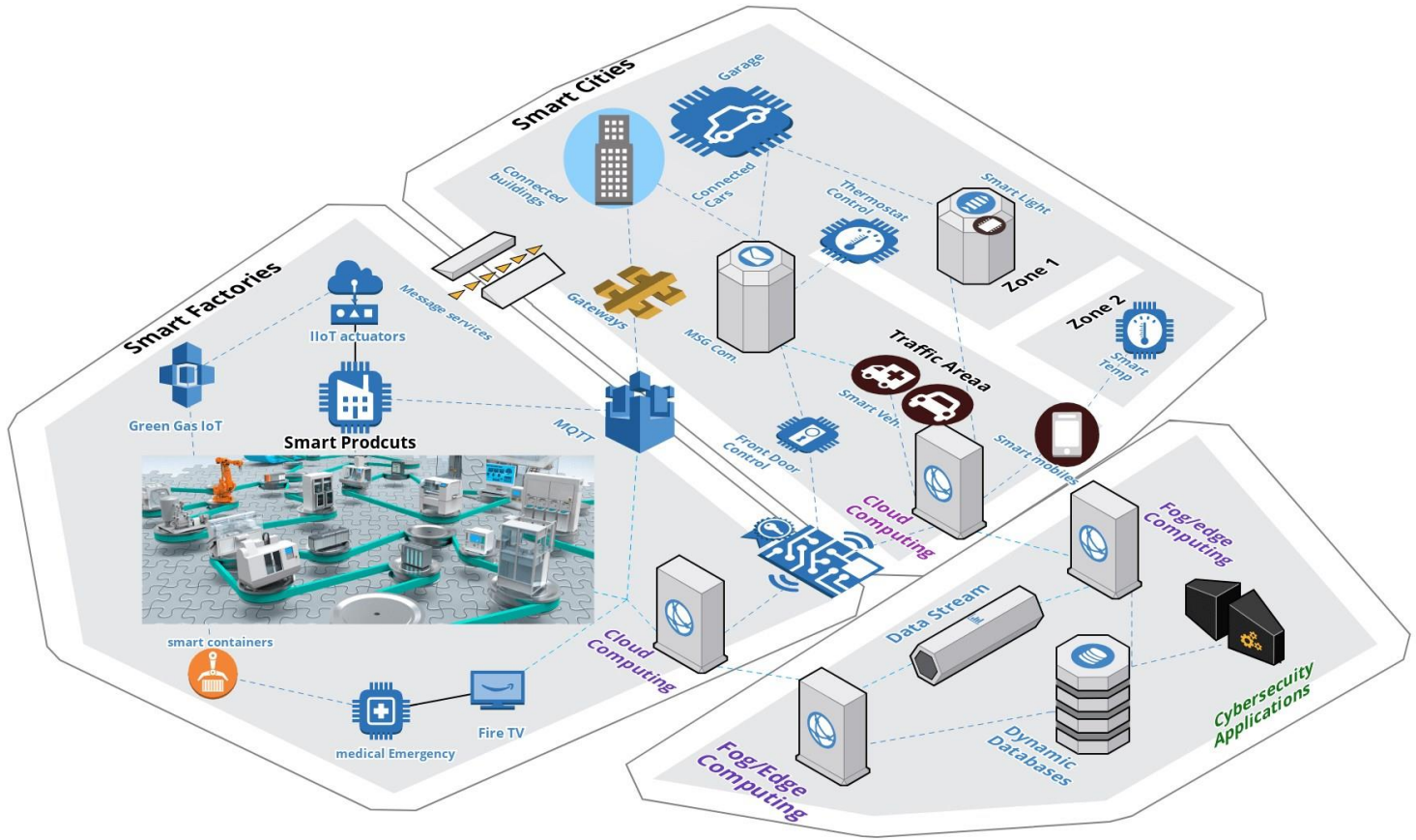

Fig. 1. Architecture of fog computing and its interaction with cloud and IoT

An example of the technical Fog architecture was proposed by Cisco is shown in Figure 2 [1] to design the fog architecture as IaaS. The Cisco IOx platform operates by hosting programs in an operating system that runs a hypervisor on a grid router. The IOx APIs allow the Fog to connect with IoT and Cloud systems by a user-identified protocol. For designing the Fog as cyber security applications, the architecture could be breached by sophisticated cyber-attacks, such as Distributed Denial of Service (DDoS) and ransomware, because the Fog nodes could be connected with unsecured and non-standard IoT sensors-based IP addresses. Therefore, different security systems should be deployed at the Fog nodes to mitigate the risk impacts of cyber threats.

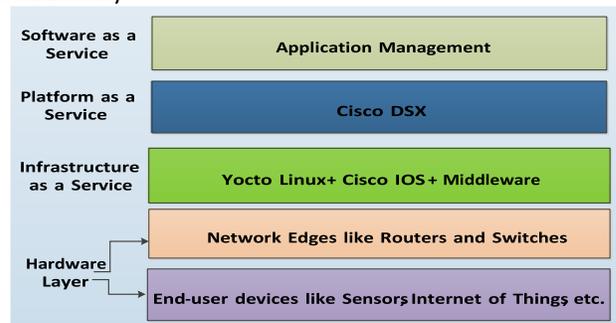

Fig. 2. Cisco's Fog technical architecture

*D. Applications of IoT, Fog and Cloud Systems*

The advantages of the Fog can be applied to different IoT and Cloud systems [4], [26]. This demonstrates how the Fog paradigms can be implemented in real-time and large-scale systems, as explained in the following applications:

- Smart grid: includes smart meters and micro-grids implemented at the edge of a network as energy load balancing services. The Fog can support processing smart grid nodes at the network edges. Data generated from IoT networks are stored at Fog nodes for running big-data analytics and cyber security applications [28].
- Software Defined Networks (SDN): is a promising computing and network architecture. The Fog can be used for designing a SDN architecture to manage and control the SDN communication layers. The control unit is executed at a centralised server, where the nodes of SDN can execute a communication path specified by a server which requires distributed executions [26].
- Linked vehicles and smart traffic systems: are improved by the connection with the Fog nodes, such as a vehicle to access points or vice versa. The smart traffic systems interact with different sensors at the network edge to send warning lights to the coming cars for avoiding possible accidents. Connecting these smart grids with the Fog could address the drawbacks of low latency, low mobility support and geographical distributions [4], [13].
- Wireless sensor and actuator networks: are used for sensing and tracking different IoT applications, with the dependency on actuators to control physical systems. When actuators operate as Fog appliances, they can easily manage the performance of systems [16].
- Industry 4.0 and Industrial IoT systems: Industry 4.0 systems include the applications of cyber-physical systems, Industrial IoT (IIoT) and IoT. The aim of these systems is to link physical devices to the Internet and Cloud systems. These systems can be used for rapidly processing and storing different heterogeneous sources at the network edges and improving security issues [29], [30].

III. CYBER SECURITY CHALLENGES

Since Fog devices are connected with the Cloud and IoT systems, IoT networks could be exploited using different cyber threats. This is because the devices are deployed at unsecured locations which are not accurately monitored and protected. The open architecture of the Fog leads to loopholes and vulnerabilities that allow attackers to compromise the Fog devices and services, in addition to threaten the privacy of its big-data [35]. Different security issues could face the Fog-IoT-Cloud architecture, as discussed below.

- Authentication and authorisation - Fog devices could be connected with the Cloud servers via a distributing authentication system, but this connection is relatively slow in smart grids [28]. The execution of authentication protocols, for example, directory access and remote authentication, are improper due to the limitation of connections. Moreover, using Cloud servers for authentication is not a right solution as they would be penetrated by bruteforce and dictionary attacks for stealing user credentials [4].

TABLE II
ATTACK TYPES THAT COULD EXPLOIT ELEMENTS OF IOT-FOG-CLOUD ARCHITECTURE

| Attack types | Description |
| --- | --- |
| Insider intruders | refer to authorised cloud users who attempt to gain unauthorized rights, penetrating cloud resources with no privileges |
| Attacks on Virtual Machines (VM) or hypervisor | when the virtual layer of hypervisor is compromised using zero-day attacks, attackers can control the installed VMs and physical hosts |
| Flooding attacks | An attacker attempts to flood a victim by sending a lot of packets via DoS and DDoS from a computer host in a network (i.e., zombie) to breach VMs |
| Service abuses | can be hijacked by malicious activities, for example, using cloud/fog computing resources to violate an encryption key to launch an attack |
| Advanced persistent threats (APTs) | penetrate systems to launch a footprint attack, then stealthily infiltrate data and intellectual property continually |
| Port scanning | finds a list of all open ports, closed ports and filtered ports in a network. Attackers searching for finding open ports to get access to a particular system |
| Backdoor attacks | are passive attacks in which an hacker bypasses a stealthy normal authentication mechanism to protect unauthorised remote access to a device. an attacker could control victim's resources and make it as a zombie to initiate DoS/DDoS attacks |
| User to Root (U2R) attacks | an attacker gains an access to legitimate user's account by sniffing password. This leads to breach exposures for gaining the root level access to victim's device, e.g., Buffer overflows |

- Advanced Persistent threats- Fog systems face various sophisticated attack types, such as botnets and ransomware, inherited from Cloud and IoT systems. These cyber attacks would expose Fog nodes, due to its distributed architecture [1], [4], [5], [13], as summarised in Table II.
- Suspicious Fog nodes- since Fog nodes handle big-data collected from IoT devices, dividing workloads between the

nodes is often heavy. In this sense, if an attacker compromises any of the nodes, it is hard to assert data integrity and privacy. Trust mechanisms should be deployed to ensure data transfer between Fog and Cloud systems [32].

- Fog data management- since Fog nodes are geographically distributed, it is difficult to know the location of data gathered from Cloud systems. It is hard for customers to identify either the node offers the same service or not [36]. Some Fog nodes often contain duplicated data with other nodes that consume resources, and it is possible that attack events are injected to this data using data poisoning techniques.

Fog-Cloud architecture. Each security tool can be utilised for handling a specific security challenge, described in Table III and explained in the following:

- Authentication technique- is the process of identifying users with different methods. Fog computing should include biometric authentication that involves face, fingerprint, balm, touch-based or keystroke-based methods. They are promising solutions compared with traditional methods, such as password-based authentication [31]. In [4], the authors stated that one of the key security challenges for Fog computing is authentication mechanisms at various levels of Fog nodes

TABLE III
THREATS, ADVANTAGES AND DISADVANTAGE OF EXISTING SECURITY SOLUTIONS

| Solutions | Threats | Advantages | Disadvantages |
|---|---|---|---|
| Authentication techniques [31] | insider attacks, including brute force and dictionary attacks | - are easy to use and reduce operational costs<br>- enhance customer experience | - cannot be reset once exploited<br>- demand integration with different Fog devices |
| Access control systems [32] | birthday, sniffer, spoofing and phishing attacks | - are capable of achieving accessibility and optimal control using several options like biometrics and federated identify keys<br>- are easy to integrate with other security controls and mange their database | - are expensive to install, as they include an upfront financial investment<br>- demand regular updates to reduce the chances of hacking |
| Intrusion detection systems [33], [34] | insider attacks, flooding, VM attacks, APTs, U2R attacks, backdoor and port scanning attacks | - can be adapted to a particular content in a network for boosting the efficiency<br>- make it easier to continue with regulation | - do not process encrypted packets and handle header packets only -<br>produce high false alarm rates |
| Privacy and encryption techniques [32] | flooding attacks and service abuses | - improve security, as private keys do not transmit over networks - can offer a mechanism for digital signatures | - when attackers collect enough information, they can violate keys<br>- the key methods have to be regularly updated |

- Privacy issues - deal with concealing confidential information, such as what device were used in a particular time while enabling data summarisations to be exchanged between fog nodes. Privacy preservation techniques should hide details of sensitive information about Fog devices and services, for example, what devices are used at a certain time. Existing Fog appliances cannot encrypt and decrypt the readings of smart meters. Therefore, those appliances could expose sensitive information while transmitting and receiving data flows across network nodes [4], [37].

IV. SECURITY SOLUTIONS AND FUTURE DIRECTIONS

Various security solutions have been employed, for example, authentication, access control, encryption, firewall, as well as intrusion detection and prevention systems, for addressing different security and privacy challenges at the IoT-using public key techniques. In [32], [38], the trusted execution mechanism should have its potential in Fog computing in order to decrease the complexity of authentication.

- Access control- is a trustworthy mechanism installed at IoT and Cloud devices that guarantee authentication and authorisation to end-users and workstations, along with servers [32]. In Fog , a policy-based control was proposed to protect the cooperation between heterogeneous sources [39]. There is still a challenge of how to design an effective access control system for clients in IoT networks to protect systems at different levels.

- Intrusion detection system (IDS) - can be installed in the Fog layer to recognise suspicious events by inspecting audit traces of the client side. It can also be installed at the fog side to identify suspicious attacks by analysing network traffic [33], [34] . In [40], the authors suggested a cloudlet mesh based on a security framework that can identify attacks from Cloud and Fog systems. There are still the challenges of implementing scalable and adaptive intrusion detection at the fog layer to achieve the lowlatency requirements [32].

- Privacy and encryption techniques- protecting user information is one the biggest issues in IoT, Fog and Cloud systems. Various privacy-preserving mechanisms have been suggested in the Cloud, smart grids and wireless networks. These mechanisms could be implemented between the Cloud and Fog layers to prohibit tampering big-data transmitted between the two layers. Encryption techniques should be applied to obfuscate data exchange between different network nodes [32]. However, because of the distributions of network nodes, privacy techniques need further research for protecting sensitive information of users.

## V. Conclusion

In this paper, an architecture has been proposed to illustrate the interactions of IoT, Cloud and Fog layers for effectively running big-data anaytics and cyber security applications. Since the devices and services in the three layers generate heterogeneous data sources, the Cloud systems have been used to process, compute and store such data at centralised locations. However, the mobility-support, location-awareness, low latency and geographical location are still the key challenges in the Cloud layer that could be tackled using the Fog paradigms by processing computational tasks at the edge of the network. The use of fog technology still faces security and privacy challenges that originate from the connection with the open architecture of IoT and Cloud systems. The security problems in existing security tools and future research directions are introduced to improve the security of the IoTFog-Cloud architecture.

## References

[1] S. Khan, S. Parkinson, and Y. Qin, "Fog computing security: a review of current applications and security solutions," *Journal of Cloud Computing*, vol. 6, no. 1, p. 19, 2017.

[2] N. Moustafa, K.-K. R. Choo, I. Radwan, and S. Camtepe, "Outlier dirichlet mixture mechanism: Adversarial statistical learning for anomaly detection in the fog," *IEEE Transactions on Information Forensics and Security*, 2019.

[3] A. V. Dastjerdi, H. Gupta, R. N. Calheiros, S. K. Ghosh, and R. Buyya, "Fog computing: Principles, architectures, and applications," *arXiv preprint arXiv:1601.02752*, 2016.

[4] I. Stojmenovic and S. Wen, "The fog computing paradigm: Scenarios and security issues," in *Computer Science and Information Systems (FedCSIS), 2014 Federated Conference on*. IEEE, 2014, pp. 1–8.

[5] G. Kurikala, K. G. Gupta, and A. Swapna, "Fog computing: Implementation of security and privacy to comprehensive approach for avoiding knowledge thieving attack exploitation decoy technology," 2017.

[6] S. Yi, C. Li, and Q. Li, "A survey of fog computing: concepts, applications and issues," in *Proceedings of the 2015 Workshop on Mobile Big Data*. ACM, 2015, pp. 37–42.

[7] M. Chiang, S. Ha, I. Chih-Lin, F. Risso, and T. Zhang, "Clarifying fog computing and networking: 10 questions and answers," *IEEE Communications Magazine*, vol. 55, no. 4, pp. 18–20, 2017.

[8] Y. Guan, J. Shao, G. Wei, and M. Xie, "Data security and privacy in fog computing," *IEEE Network*, 2018.

[9] G. Premsankar, M. Di Francesco, and T. Taleb, "Edge computing for the internet of things: A case study," *IEEE Internet of Things Journal*, 2018.

[10] R. Roman, J. Lopez, and M. Mambo, "Mobile edge computing, fog et al.: A survey and analysis of security threats and challenges," *Future Generation Computer Systems*, 2016.

[11] K.-K. R. Choo, R. Lu, L. Chen, and X. Yi, "A foggy research future: Advances and future opportunities in fog computing research," 2018.

[12] N. Moustafa, G. Misra, and J. Slay, "Generalized outlier gaussian mixture technique based on automated association features for simulating and detecting web application attacks," *IEEE Transactions on Sustainable Computing*, 2018.

[13] M. Aazam and E.-N. Huh, "Fog computing and smart gateway based communication for cloud of things," in *Future Internet of Things and Cloud (FiCloud), 2014 International Conference on*. IEEE, 2014, pp. 464–470.

[14] A. Alrawais, A. Alhothaily, C. Hu, and X. Cheng, "Fog computing for the internet of things: Security and privacy issues," *IEEE Internet Computing*, vol. 21, no. 2, pp. 34–42, 2017.

[15] F. Y. Okay and S. Ozdemir, "A fog computing based smart grid model," in *Networks, Computers and Communications (ISNCC), 2016 International Symposium on*. IEEE, 2016, pp. 1–6.

[16] Q. Yaseen, F. AlBalas, Y. Jararweh, and M. Al-Ayyoub, "A fog computing based system for selective forwarding detection in mobile wireless sensor networks," in *Foundations and Applications of Self* Systems, IEEE International Workshops on*. IEEE, 2016, pp. 256–262.

[17] N. I. M. Enzai and M. Tang, "A taxonomy of computation offloading in mobile cloud computing," in *Mobile Cloud Computing, Services, and Engineering (MobileCloud), 2014 2nd IEEE International Conference on*. IEEE, 2014, pp. 19–28.

[18] V. K. Sehgal, A. Patrick, A. Soni, and L. Rajput, "Smart human security framework using internet of things, cloud and fog computing," in *Intelligent distributed computing*. Springer, 2015, pp. 251–263.

[19] R. Sandhu, A. S. Sohal, and S. K. Sood, "Identification of malicious edge devices in fog computing environments," *Information Security Journal: A Global Perspective*, pp. 1–16, 2017.

[20] "Open fog consortium," Aug. 2018. [Online]. Available: http://www.openfogconsortium.org/,

[21] S. K. Datta, C. Bonnet, and J. Haerri, "Fog computing architecture to enable consumer centric internet of things services," in *Consumer Electronics (ISCE), 2015 IEEE International Symposium on*. IEEE, 2015, pp. 1–2.

[22] M. Zhanikeev, "A cloud 19 platform to facilitate cloud federation and fog computing," *Computer*, vol. 48, no. 5, pp. 80–83, 2015.

[23] W. S. Chin, H.-s. Kim, Y. J. Heo, and J. W. Jang, "A context-based future network infrastructure for iot services," *Procedia Computer Science*, vol. 56, pp. 266–270, 2015.

[24] S. Yi, C. Li, and Q. Li, "A survey of fog computing: concepts, applications and issues," in *Proceedings of the 2015 Workshop on Mobile Big Data*. ACM, 2015, pp. 37–42.

[25] H. T. Dinh, C. Lee, D. Niyato, and P. Wang, "A survey of mobile cloud computing: architecture, applications, and approaches," *Wireless communications and mobile computing*, vol. 13, no. 18, pp. 1587–1611, 2013.

[26] "Etsi:Mobile-edge computing," 2014. [Online]. Available: http://goo.gl/7NwTLE

[27] M. Mukherjee, R. Matam, L. Shu, L. Maglaras, M. A. Ferrag, N. Choudhury, and V. Kumar, "Security and privacy in fog computing: Challenges," *IEEE Access*, vol. 5, pp. 19293–19304, 2017.

[28] "Nist. guidelines for smart grid cyber security (nist 7628),," Aug. 2018. [Online]. Available: Http://csrc.nist.gov/publications/ PubsNISTIRs.html

[29] G. Peralta, M. Iglesias-Urkia, M. Barcelo, R. Gomez, A. Moran, and

J. Bilbao, "Fog computing based efficient iot scheme for the industry 4.0," in *Electronics, Control, Measurement, Signals and their Application to Mechatronics (ECMSM), 2017 IEEE International Workshop of*. IEEE, 2017, pp. 1–6.

[30] N. Moustafa, E. Adi, B. Turnbull, and J. Hu, "A new threat intelligence scheme for safeguarding industry 4.0 systems," *IEEE Access*, vol. 6, pp. 32910–32924, 2018.

[31] F. Bonomi, R. Milito, J. Zhu, and S. Addepalli, "Fog computing and its role in the internet of things," in *Proceedings of the first edition of the MCC workshop on Mobile cloud computing*. ACM, 2012, pp. 13–16.

[32] K. Lee, D. Kim, D. Ha, U. Rajput, and H. Oh, "On security and privacy issues of fog computing supported internet of things environment," in *Network of the Future (NOF), 2015 6th International Conference on the*. IEEE, 2015, pp. 1–3.

[33] N. Moustafa, J. Slay, and G. Creech, "Novel geometric area analysis technique for anomaly detection using trapezoidal area estimation on large-scale networks," *IEEE Transactions on Big Data*, 2017.

[34] N. Moustafa, G. Creech, and J. Slay, "Big data analytics for intrusion detection system: Statistical decision-making using finite dirichlet mixture models," in *Data Analytics and Decision Support for Cybersecurity*. Springer, 2017, pp. 127–156.

[35] H. Dubey, J. Yang, N. Constant, A. M. Amiri, Q. Yang, and K. Makodiya, "Fog data: Enhancing telehealth big data through fog computing," in *Proceedings of the ASE BigData & SocialInformatics 2015*. ACM, 2015, p. 14.

[36] A. Sinaeepourfard, J. Garcia, X. Masip-Bruin, and E. Marin-Tordera, "A novel architecture for efficient fog to cloud data management in smart cities," in *Distributed Computing Systems (ICDCS), 2017 IEEE 37th International Conference on*. IEEE, 2017, pp. 2622–2623.

[37] M. Keshk, E. Sitnikova, N. Moustafa, J. Hu, and I. Khalil, "An integrated framework for privacy-preserving based anomaly detection for cyberphysical systems," *IEEE Transactions on Sustainable Computing*, 2019.

[38] C. Chen, H. Raj, S. Saroiu, and A. Wolman, "ctpm: A cloud tpm for cross-device trusted applications." in *NSDI*, 2014, pp. 187–201.

[39] C. Dsouza, G.-J. Ahn, and M. Taguinod, "Policy-driven security management for fog computing: Preliminary framework and a case study," in *Information Reuse and Integration (IRI), 2014 IEEE 15th International Conference on*. IEEE, 2014, pp. 16–23.

[40] C. Modi, D. Patel, B. Borisaniya, H. Patel, A. Patel, and M. Rajarajan, "A survey of intrusion detection techniques in cloud," *Journal of Network and Computer Applications*, vol. 36, no. 1, pp. 42–57, 2013.